\documentstyle[preprint,aps]{revtex}
\begin{document}
\draft

\title{
{\it Ab initio} Molecular Dynamics study of electronic and optical
properties of silicon quantum wires: Orientational Effects
}

\author{A. M. Saitta$^{a}$, F. Buda$^{b}$, G. Fiumara$^{c}, $ P. V.
Giaquinta$^{d}$}

\address{
$^a$ Istituto Nazionale per la Fisica della Materia,
Scuola Internazionale Superiore di Studi Avanzati (SISSA),
Via Beirut 2-4, I-34014 Trieste, Italy }

\address{
$^b$ Istituto Nazionale per la Fisica della Materia, Laboratorio Forum,
Scuola Normale Superiore,
Piazza dei Cavalieri 7, I-56126 Pisa, Italy}

\address{
$^c$  Istituto Nazionale per la Fisica della Materia,
Unit\`a di Ricerca di Messina,
CP 50 I-98166 S. Agata, Messina, Italy}

\address{
$^d$  Istituto Nazionale per la Fisica della Materia and Universit\`a
degli Studi di Messina, Dipartimento di Fisica, Sezione
di Fisica Teorica,
CP 50 I-98166 S. Agata, Messina, Italy}

\date{\today}
\maketitle
\pagebreak
\begin{abstract}

We analyze the influence of spatial orientation
on the optical response of hydrogenated silicon quantum wires.
The results are relevant for the interpretation of the optical
properties of light emitting porous silicon.
We study (111)-oriented wires and compare the present
results with those previously obtained within the same theoretical framework
for (001)-oriented wires [F. Buda, J. Kohanoff and M. Parrinello, {\it
Phys. Rev. Lett.} {\bf 69}, 1272, (1992)].
In analogy with the (001)-oriented wires and at variance with crystalline
bulk silicon, we find that the (111)-oriented wires exhibit a direct gap
at ${\bf k}=0$ whose value is largely enhanced with respect to that found in
bulk
silicon because of quantum confinement effects.
The imaginary part of the dielectric function, for the external field
polarized in the direction of the axis of the wires, shows features that,
while being qualitatively similar to those observed for the (001) wires,
are not present in the bulk.
The main conclusion which emerges from the present study is that, if wires
a few nanometers large are present in the porous material, they are
 optically active independently of their specific orientation.

\end{abstract}

\narrowtext

\vskip 0.4truecm
\section{Introduction}
\medskip

Silicon, one of the most relevant materials from the technological point
of view, has excellent electronic properties as
a semiconductor. However, in the crystalline form, it does not exhibit
good optoelectronic properties because of the indirect gap, in the
infrared range, between the valence and the conduction bands, and of the
low probability of radiative electron-hole recombination.
However it is possible to modify deeply such properties using a process
of electrochemical etching~\cite{uhl} which produces a material
known as light emitting porous silicon (LEPS).
This form, because of its luminescence properties,
is presently the object of great scientific and technological
attention~\cite{can}.
In fact, inside this material one can observe some wire structures:
the wires are strictly interconnected with each other,
and have a mean diameter of a few tens of \AA ngstrom.
Quantum confinement of charge carriers seems to be at the origin of the
optical characteristics of the material, which are quite different from those
of bulk
silicon.

In the recent past many different theoretical
studies~\cite{bkp,PQE,hyn,rea,ohn,del,hyb,bis} have been developed
with the aim of analyzing the actual validity of the quantum confinement
hypothesis
in order to explain those new optical effects.
According to the current experimental evidence, besides the presence of
nanostructures of reduced dimensionality (wires and/or dots), the key
ingredients
that are necessary to induce luminescence are crystallinity and surface
passivation
of dangling bonds.
In the present work we concentrate on the quantum wire model proposed to
explain
the optical properties of LEPS. We use the {\it ab initio} Molecular Dynamics
(AIMD)
method, as proposed by Car and Parrinello~\cite{par,gal,pas}, to study silicon
quantum wires oriented along the (111) crystallographic direction.
The AIMD method has been successfully applied to a variety of different
systems and problems~\cite{gal}.
Porous silicon samples grown on a silicon substrate in the (111) lattice
direction
were characterized
both in the morphology and in the optical properties~\cite{mas1,mas2,per}.
The study presented here follows a similar work developed for quantum wires
oriented along the (100) crystallographic direction~\cite{bkp}.
This study showed the presence of a direct gap in the visible range
and an important enhancement of the dipole matrix elements, which are
directly related to the transition probability between the valence and the
conduction band and, in consequence of that, to the absorption coefficient.
This work, such as all the others quoted above, shows how the optical
and electronic properties of silicon become deeply modified by quantum
confinement.
These predictions, with the limitations due to the Local Density Approximation
(LDA) used in the AIMD framework,
support the thesis that the origin of porous silicon
photoluminescence lies in the localization of charge carriers inside ordered
structures of reduced dimensionality.

The present work tries mainly to ascertain if the orientation may be relevant
in determining the optical
and electronic properties of the material since, {\it a priori}, it is not
obvious
that charge carriers confinement acts in a similar way along different
symmetry planes.
In section II we describe the geometry of the model used and give some
computational details. In section III the results on the structural,
vibrational and electronic properties of the model are discussed.
The final section is devoted to the conclusions.

\vskip 0.4truecm
\section{Silicon wire geometry and technical details}
\medskip

Crystalline silicon is characterized by a diamond structure, in which atoms
are distributed in a face centered cubic (fcc) lattice, each lattice site being
occupied by a pair of particles whose chemical bond, about 2.35 \AA ~long,
lies in the (111) direction.
The silicon lattice structure along the (111) direction can be characterized by
three double layers periodically repeated with a periodicity
$a{\sqrt 3}$, where $a$ is the lattice parameter of bulk crystalline
silicon (5.43 \AA).
Fig.~\ref{f:struc} shows the smallest wire that one can build,
still preserving the crystalline order along the (111) direction.
The unit cell, indefinitely replied in space
through periodic boundary conditions (PBC), contains 26 silicon atoms, plus
30 hydrogen atoms which are necessary to saturate the dangling bonds at the
surface of the wire.
The geometry of the model is such that it exhibits a non uniform size along the
longitudinal direction. In a way, this feature makes the model more similar to
real
samples of porous silicon.
In fact, the diameter of the model has a maximum value of
7.67 \AA, and a mean value of 5.51 \AA.
We have also studied a larger wire whose basic
structure consists of 62 silicon atoms and 42 hydrogen atoms,
with a maximum diameter of 11.72 \AA, and a mean value of 10.37 \AA.
We note that a system consisting of 104 atoms is rather demanding
on the computational side for {\it ab-initio} Car-Parrinello simulations.
For this reason, we limited our calculations for the larger wire to those
features
which may confirm some specific properties of the thinner wire.
In both cases the size of the supercell in the longitudinal direction
corresponds exactly to the bulk silicon characteristic size of 9.41 {\AA}
along the (111) direction.
The lateral size of the supercell is much larger than
the diameter of the wire so as to avoid interactions
of the particles with their images.
The parameters needed in the AIMD simulation were suggested by the
previous experience on this material:
we used a time step of 7 a.u., an energy cutoff
for the plane-wave expansion of 8 Ry, an electronic mass of 300 a.u.,
and pseudopotentials of the Bachelet-Hamann-Schluter~\cite{bac} type for
describing the interaction between valence electrons and the nuclei
with core electrons.

\vskip 0.4truecm
\section{Results}
\medskip

\subsection{Structural relaxation and vibrational properties}

As a first step, we wanted to verify the structural stability of the wire.
To this aim, after the electronic ground-state relative
to the initial ionic configuration was found,
we let all the ionic degrees of freedom relax with a steepest descent
technique.
We notice a small increase of the distance of the most
external atoms from the axis of the wire, of the order of 0.16 \AA,
analogously to what was observed in~\cite{bkp}. As expected,
we observe no relevant variations of the characteristic
distances, along the longitudinal direction.
Then, we let the system evolve dynamically for a time of about
2 ps, a considerable time in AIMD, so that one may be confident that the system
is reasonably close to thermal equilibrium.
During this evolution the model did not exhibit any
pathological behaviour: the system oscillates about the equilibrium
configuration with no substantial atomic rearrangements.
We observe that the total energy (potential + kinetic)
fluctuates around an average value which remains constant, as should be in a
microcanonical AIMD simulation.
The average temperature during this dynamical evolution is about 70 K.

We also computed the velocity
autocorrelation function and, through its Fourier transform, an indicative
vibrational density of states (VDOS) of the system.
We find a good
qualitative agreement with the typical experimental data of the vibrational
modes of Si-Si and Si-H bonds~\cite{tis}.
 From a strictly quantitative point of view, we remark that the
relative intensity of the peaks should not be taken too rigorously.
However the positions of the peaks are in fair
agreement with the experimental data.
In particular we recognize in the spectra of Fig.~\ref{f:vibra}
the characteristic signal of the stretching of the Si-H bond
at about $2000$ $cm^{-1}$. The double peak
has its origin in the presence of doubly hydrogenated
silicon atoms, which give rise to symmetrical and antisymmetrical
stretching of the bond at slightly different frequencies.
The analysis of the oscillations of single and twin hydrogen particles
shows that the higher-frequency peak can be ascribed to the symmetrical mode,
while the lower-frequency signal arises from antisymmetrical stretching
vibrations.
We notice also, at lower frequencies, the typical signals of scissors mode
($~800$ $cm^{-1}$) of the twin hydrogen atoms bonded to one silicon atom,
and of Si-H bending (around $~600$ $cm^{-1}$).
The weak structure centered at about $1150$ $cm^{-1}$ is likely associated
with another bending vibrational mode of hydrogen atoms as it shows up
in the motion of both single and twin passivating hydrogens.
The low frequency noise produced by the finite time evolution renders
more uncertain the identification of the modes below $500$ $cm^{-1}$ that are
associated with the motion of silicon atoms.
It is important to remark that such characteristic vibrational modes
are observed, by infrared spectroscopy, also in real freshly prepared
LEPS samples~\cite{tis}.

For the larger size sample, we have also studied the relaxation
towards the global energy minimum. It is noteworthy
that in this case the mean radial relaxation, in the equilibrium
configuration, is negligibly small if compared with that estimated
for the thinner wire.
This can be intuitively understood if one thinks that in the
smaller-size model the ratio of surface to bulk atoms is definitely
higher than in the larger wire.
Such a small relaxation originates from the competition between the
doubly hydrogenated silicon atoms, which move slightly away from the axis of
the wire, and the singly hydrogenated ones, which instead move inward.

\subsection{Electronic and optical properties}

Once the minimum energy configuration has been determined, we computed
the electronic band
structure of the quantum wire along the one-dimensional Brillouin zone (BZ).
We computed the Kohn-Sham eigenvalues for six different k-points of the BZ,
from the center ($\Gamma$) to the border (which would correspond to L in the
bulk).
The resulting band structure is shown
in Fig.~\ref{f:band} for the (111)-oriented wire with a mean
diameter of 5.51 \AA.
The main result is the prediction of a direct gap of 2.83 eV between the
conduction and the valence band at the zone center.
It is also interesting to notice that the second conduction state
still shows a signature of the band structure of bulk silicon,
i.e. a minimum located at about 0.2.
As to the gap, we recall that Buda and co-workers found, for a (100)-oriented
quantum wire with a diameter of 7.60 \AA, a value of 2.33 eV~\cite{bkp}.
The difference is to be ascribed, as we shall see later,
to the different mean diameter of the two models,
more than to the different symmetry of the wires.
We recall that in bulk silicon the minimum of the conduction band along
the $\Gamma$-L direction falls just at L, with a gap of 2.15 eV.

For the larger wire we find an energy gap of 1.90 eV at the center of BZ
which, consistently with the quantum confinement
hypothesis, is larger than that in bulk silicon, but
smaller than the value for the thinner wire.
The values of the two energy gaps, together with those of the (100) models
and of bulk silicon as obtained in LDA, are plotted in
Fig.~\ref{f:egap}.
It is of striking evidence the linear behaviour of the energy gap as a
function of the inverse size of the models. It is remarkable that both
the (111) and (100) computed values lie on the same straight line,
suggesting the independence of such property on the specific
orientation of the wires.
In the same figure we also show the theoretical prediction coming from
effective mass arguments (E$_g \sim {1 \over d^{2}}$),
which should hold close to the bulk limit.
It appears that the matching between these two different behaviours
should occur in the region between 2 and 5 nm.

The presence of a direct gap, and thus the possibility
of zero-phonon transitions, is not a sufficient condition for ensuring good
photoluminescence properties. The probability of an electronic transition
as a result of the absorption of a photon is given by the Fermi golden rule,
which in the dipole approximation is
$|\langle \psi_{c{\bf k}}|{\bf e} \cdot {\bf p}|\psi_{v{\bf k}}\rangle|^2$
where $\psi_{c{\bf k}}$ and $\psi_{v{\bf k}}$ are the conduction and valence
states involved in the electronic transition, while
{\bf e} is the polarization vector and {\bf p} the momentum operator.
We determined all the matrix elements between Kohn-Sham states of the
conduction
and valence bands whose difference in energy falls in the visible
range, or slightly above.
Table~\ref{t:small}  and Table~\ref{t:big}
give the values of the most relevant of these dipole-matrix
elements for the smaller and the larger (111)-oriented wires.
We note the strong difference between the values relative to the
longitudinal and transverse polarizations. Moreover, the quantum confinement
of the electrons mixes up the states, thus allowing for transitions
that are forbidden in the bulk by selection rules.
The dipole-matrix element between the
top of the valence band and the bottom of the conduction band is
not zero because of the mixing, even though not very large.

We computed the imaginary part of the dielectric function, that
is related to the absorption coefficient, through the formula
\begin {equation}
\epsilon_2(\omega) = {{4\pi^2 e^2} \over {m^2 \omega^2}}{{L_w} \over {V}}
                      \sum_{v,c}\int_{BZ}{{2dk_z} \over {2\pi}}
      \left|\langle \psi_{ck_z}|{\bf e} \cdot {\bf p}|\psi_{vk_z}\rangle
\right|^2
                      \delta(E_c(k_z)-E_v(k_z)-\hbar\omega)
\end {equation}
where $L_w$ and $V$ are the length of the wire and the volume of the supercell,
respectively.
The results for the smaller model are shown in Fig.~\ref{f:eps2small}.
In the construction of this quantity we have considered only the dipole-matrix
elements corresponding to energy differencies up to 4 eV.
We observe that the component related to
an external field parallel to the wire direction
is much stronger than the perpendicular one, and
this difference becomes larger and larger going to lower energies.
Probably, the most interesting result is the presence, in the parallel
component,
of a small signal at the band gap (2.83 eV) which is completely absent
in the perpendicular one.
This is a clear effect of the mixing of the bulk states induced by the
reduced dimensionality.
This result seems to be peculiar of the (111) wire, since in the
(100) models~\cite{bkp} the dipole-matrix element between the top
of the valence band and the bottom of the conduction band was negligible
in both the components.

The plot of the dielectric function for the larger-sized wire, that is shown
in Fig.~\ref{f:eps2big}, exhibits also a different
behaviour in the parallel and the perpendicular components.
We notice also in this case a clear difference in intensity between the
parallel and the perpendicular components at higher energies.
The most remarkable feature is the presence, below $\sim$ 2.7 eV,
of three peaks in the
longitudinal component that are not obseved in the transverse one.
In particular we observe again a significative signal in correspondence
of the energy gap (1.90 eV), whose intensity is lowered with
respect to the smaller wire. This is consistent with the fact that
in the bulk limit this matrix element should go to zero.

\vskip 0.4truecm
\section{Conclusions}
\medskip

We have presented an electronic structure calculation, within the LDA,
of two (111)-oriented silicon quantum wires of different size.
The comparison of the present results with those
obtained for the (001)-oriented models shows a substantial analogy of the
electronic and optical properties of silicon quantum wires oriented along
different crystallographic directions.
In particular we find a direct energy gap at the zone center whose magnitude
decreases for larger wires consistently with the quantum confinement
hypothesis.
Quite surprisingly we find that the values
of the gap for all the (111) and (001) wires studied
are excellently fitted as a linear function of the inverse diameter.
Therefore this quantity
seems to be independent on the orientation and geometrical details,
and is affected only by the width of the wire.
Again, the present calculations support the idea
that the photoluminescence properties of
porous silicon can be ascribed to charge carriers confinement
in nanostructures.
We find, similarly to the (001) wires, a strong anisotropy in
the imaginary part of the dielectric function.
These results indicate that, if the LEPS structure consists of a network
of randomly oriented wires~\cite{tsd}, all of them will contribute to
the photoluminescence independently on their specific orientation.

{\bf Acknowledgements}:
This work was supported by the Ministero dell'Universita' e della Ricerca
Scientifica e Tecnologica (MURST) and by the Istituto Nazionale per la
Fisica della Materia (INFM), Italy.
The authors thank the Centro di Calcolo Elettronico dell'Universita'
degli Studi di Messina for the allocation of computer time.
A. M. S. would also like to acknowledge the hospitality of the
IBM-ECSEC centre in Rome in the early stage of this work.
Very useful discussions with J. Kohanoff, S. Ossicini, and O. Bisi are also
acknowledged.

%
%
\begin{figure}
\caption{
Silicon quantum wire oriented in the (111) direction:
side view (panel a) and top view (panel b).
The large balls represent silicon atoms while the small balls
indicate the hydrogen atoms passivating the dangling bonds at the surface
of the wire.
}
\label{f:struc}
\end{figure}

\begin{figure}
\caption{
Vibrational spectrum for a (111)-oriented
wire with an average diameter of 5.51 \AA.
We rescaled of a factor 5 the feature at about 2000 $cm^{-1}$ in order to
emphasize the double peak (see text).
}
\label{f:vibra}
\end{figure}

\begin{figure}
\caption{
Electronic band structure for the smaller (111)-oriented wire along the
one-dimensional Brillouin zone. We express k in unit of
$\frac {2 \pi} {c}$, where c is the periodicity along the direction
of the wire.
Only few bands close to the energy gap are plotted.
}
\label{f:band}
\end{figure}

\begin{figure}
\caption{
Computed energy gap as a function of the inverse wire diameter.
We show both the (111) wires (diamonds) and (100) wires (crosses) computed
gap. The full line is the linear fit to this points. The dashed curve
represents the effective mass theory prediction. The LDA bulk silicon gap is
represented by the square on the vertical axis.
}
\label{f:egap}
\end{figure}

\begin{figure}
\caption{
Imaginary part of the dielectric function $\epsilon_2 $(E)
for the thinner wire. The plot is on a semilogarithmic scale.
The full line represents $\epsilon_2^{\parallel}$, while the dashed
line refers to $\epsilon_2^{\perp}$.
}
\label{f:eps2small}
\end{figure}

\begin{figure}
\caption{
Imaginary part of the dielectric function $\epsilon_2 $(E)
for the larger wire. The plot is on a semilogarithmic scale.
The full line represents $\epsilon_2^{\parallel}$, while the dashed
line refers to $\epsilon_2^{\perp}$.
}
\label{f:eps2big}
\end{figure}

%
%
\begin{table}
\caption{
Square modulus of the transverse and longitudinal dipole-matrix elements
for the smaller wire in unit of $Bohr^{-2}$.
The top of the valence band and the bottom of the conduction band
are labelled by indices 67 and 68 respectively.
$E_c -E_v$ is the difference between the conduction and valence Kohn-Sham
eigenvalues.
}
\begin {tabular} {|p{0.75in}|p{0.75in}|c|c|c|}
  $N_c$
& $N_v$
& $| {\langle \psi_{c,n_c}|{\bf e} \cdot {\bf p}|\psi_{v,n_v}\rangle}_{\perp}
|^2$
& $| {\langle \psi_{c,n_c}|{\bf e} \cdot {\bf
p}|\psi_{v,n_v}\rangle}_{\parallel} |^2$
& $E_c - E_v (eV)$ \\
\hline
 68 &  65 & 0.50$\cdot10^{-3}$ & 0.11$\cdot 10^{-1}$ & 3.33 \\
 68 &  66 & 1.00$\cdot10^{-3}$ & 0.14$\cdot 10^{-1}$ & 3.32 \\
 68 &  67 & 0.23$\cdot10^{-5}$ & 0.86$\cdot 10^{-3}$ & 2.83 \\
 74 &  67 & 0.42$\cdot10^{-3}$ & 0.17$\cdot 10^{-1}$ & 3.42 \\
 76 &  67 & 0.94$\cdot10^{-2}$ & 0.45$\cdot 10^{-1}$ & 3.67 \\
\end {tabular}
\label{t:small}
\end{table}

\begin{table}
\caption{
Square modulus of the transverse and longitudinal dipole-matrix elements
for the larger wire in unit of $Bohr^{-2}$.
The top of the valence band and the bottom of the conduction band
are labelled by indices 145 and 146 respectively.
$E_c -E_v$ is the difference between the conduction and valence Kohn-Sham
eigenvalues.
}
\begin {tabular} {|p{0.75in}|p{0.75in}|c|c|c|}
  $N_c$
& $N_v$
& $| {\langle \psi_{c,n_c}|{\bf e} \cdot {\bf p}|\psi_{v,n_v}\rangle}_{\perp}
|^2$
& $| {\langle \psi_{c,n_c}|{\bf e} \cdot {\bf
p}|\psi_{v,n_v}\rangle}_{\parallel} |^2$
& $E_c - E_v (eV)$ \\
\hline
147 & 138 & 0.17$\cdot10^{-4}$  & 0.12$\cdot10^{-1}$ & 2.84 \\
148 & 139 & 0.18$\cdot10^{-4}$  & 0.12$\cdot10^{-1}$ & 2.84 \\
146 & 145 & 0                   & 0.41$\cdot10^{-3}$ & 1.90 \\
156 & 145 & 0.36$\cdot10^{-5}$  & 0.14$\cdot10^{-1}$ & 2.83 \\
167 & 145 & 0.17$\cdot10^{-4}$  & 0.57$\cdot10^{-1}$ & 3.34 \\
168 & 145 & 0.11$\cdot10^{-2}$  & 0.23$\cdot10^{-1}$ & 3.36 \\
171 & 145 & 0.17$\cdot10^{-1}$  & 0.11$\cdot10^{-1}$ & 3.44 \\
\end {tabular}

\label{t:big}
\end{table}

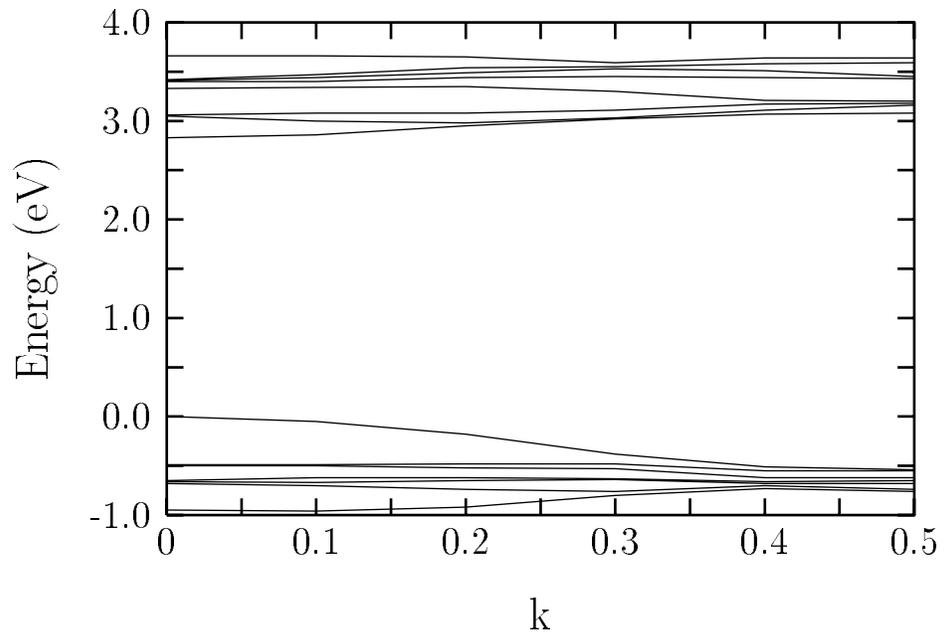
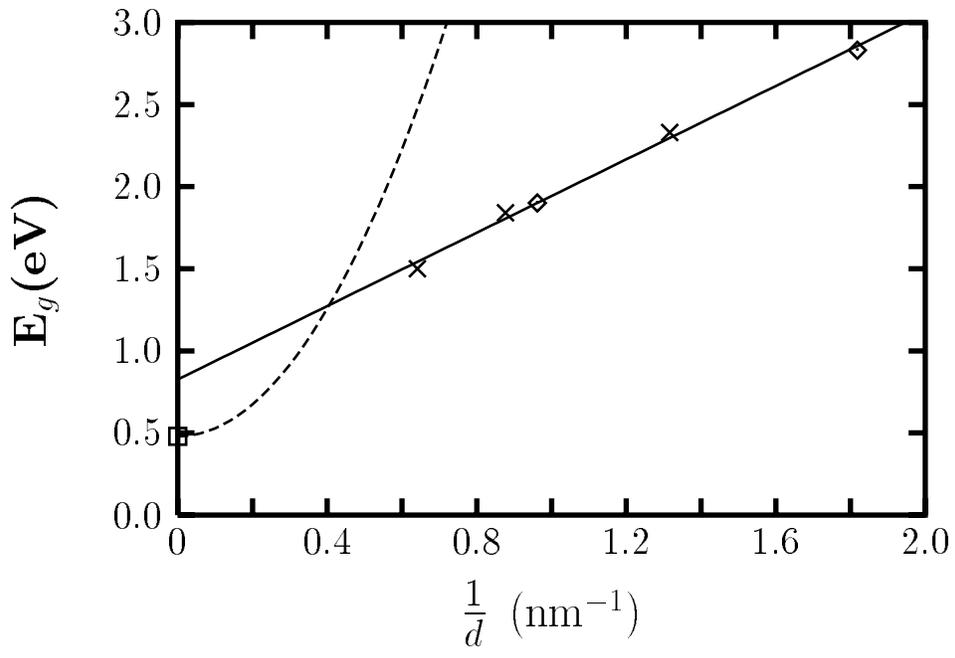
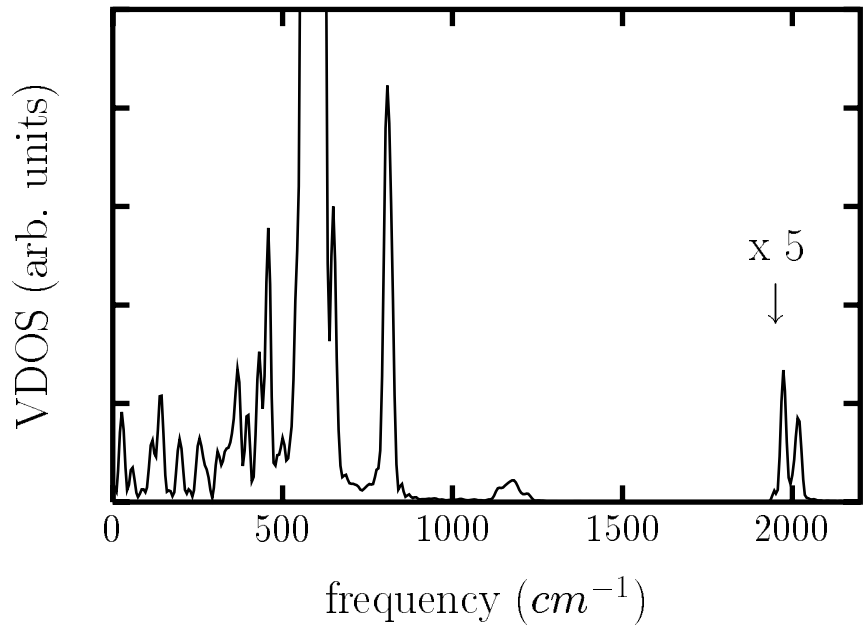
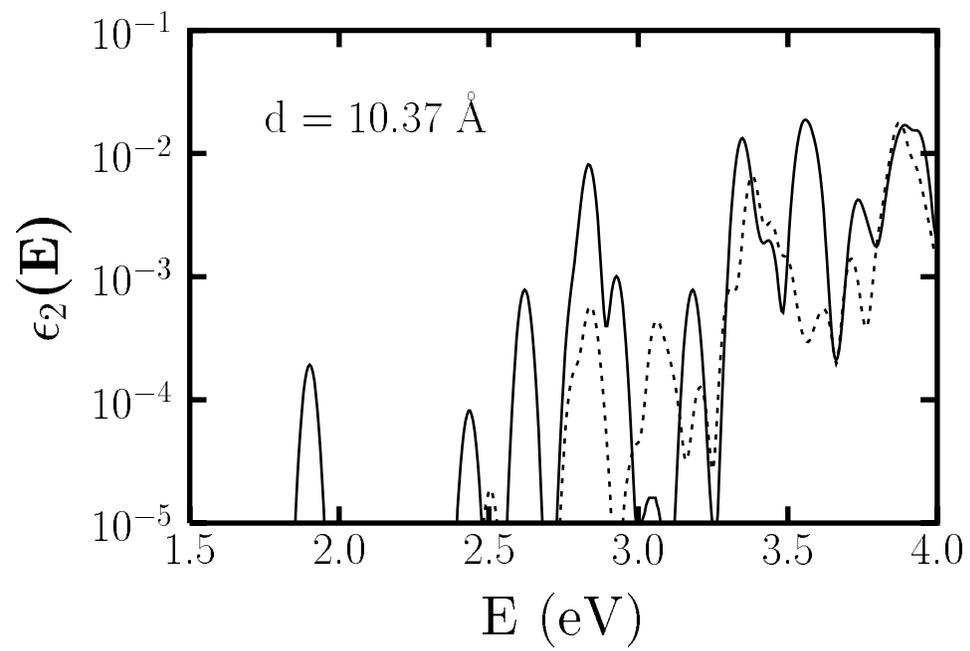
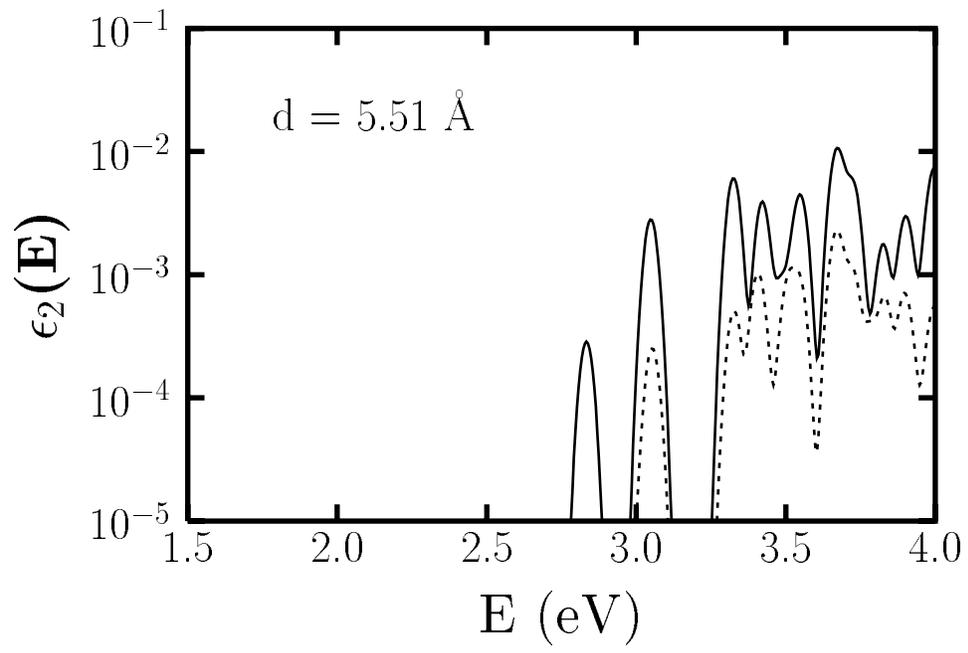
\end{document}